\documentclass[fleqn,twoside]{article}
\usepackage{espcrc2}
\usepackage{graphicx}
\usepackage[figuresright]{rotating}
\newcommand{\dmsq}{\Delta m^2}
\newcommand{\mxang}{\sin^2 (2\theta)}
\newcommand{\nmt}{\nu_\mu \rightarrow \nu_\tau}
\newcommand{\nms}{\nu_\mu \rightarrow \nu_s}
\title{Other Atmospheric Neutrino Experiments
\thanks{Contribution to the Proceedings of the XXth International
Conference on Neutrino Physics and Astrophysics, May 2002,
Munich Germany}}

\author{
Maury Goodman
\address[ANL]{HEP Division, Argonne IL 60439 USA}
}
       
\begin{document}

\begin{abstract}
The history and recent progress of atmospheric neutrinos are
reviewed.  An emphasis is placed on results from experiments 
other than 
Super-Kamiokande.
\vspace{1pc}
\end{abstract}

\maketitle

\section{Introduction}
\par The Super-Kamiokande experiment\cite{sk} has used their measurements of 
atmospheric neutrinos to persuasively and clearly
demonstrate the existence of neutrino
oscillations, with excellent agreement with the hypothesis
$\nu_\mu \rightarrow \nu_\tau$.  With lower statistics, other 
experiments have measured atmospheric neutrinos for a long time and 
continue to do so.  In this paper, after reviewing the
 creation and
measurement of atmospheric neutrinos, I discuss the
history of atmospheric neutrinos along with two possible
alternative scenarios.  I then review the most recent
results from Baksan, Soudan~2 and MACRO.
\section{The Creation of Atmospheric Neutrinos}
\par Atmospheric neutrinos originate  from the
 decays of $\pi$'s, $K$'s and $\mu$'s produced when cosmic rays 
hit the atmosphere and interact.  There are also a smaller
number of neutrinos produced by the decay of charmed particles, $\tau$'s
and other high mass particles, but their detection has not been
demonstrated.  $K$ and $\pi$  decay give mainly 
$\nu_\mu$'s while $\mu$ decay
gives both $\nu_\mu$'s and $\nu_e's$.  The $\mu$'s themselves come
from $\pi$ and K decay, so at low energy (below 2 GeV where all of the
muons decay before they hit the earth), the flux of each flavor
neutrino $\phi(\nu_\mu + \bar{\nu}_\mu):\phi(\nu_e + \bar{\nu}_e)$
occurs in the ratio 2:1.  The experimental value for this ratio (in the
absence of oscillations) for charged current interactions will vary
from 2 for several reasons:
\begin{enumerate}
\item Containment differences between $\mu$ and e
\item $\nu_\mu$ charged current threshold
\item $\nu_e$ and $\bar{\nu}_e$ differences due to their different cross 
sections and the fact that $\mu^+/\mu^-$ ratio is about 1.2
\item Some additional $\nu_e$'s from $K^+_{e3}$ and $K^0_{e3}$ decays
\item Higher energy $\mu$'s hit the earth and lose most of their energy before 
decaying.
\end{enumerate} 

\par For detectors built to study proton decay, the first factor is the
most important, and the ratio is under 2.
For high energy neutrino telescopes, the last
factor causes the ratio to increase to large values.
\par The density profile of the atmosphere affects the geometry of
the source.  The overburden, and hence the pressure,
\cite{gaisserbook} goes as
\begin{equation}
X = X_0 \exp{(-\rho g h /kT)} = X_0 \exp{(-h/h_0)}
\end{equation}
where the latter expression is used to define the scale height 
$h_0$, which  is about 8.4 km at sea level.  Since T and hence
$h_0$ depend on altitude, $h_0$ decreases to 6.4 km near the
tropopause, where many atmospheric $\nu$ are created.  Also, the air
temperature induces seasonal variations of muon and neutrino fluxes. 
Muon seasonal variations are typically $\pm 2\%$, but since high
energy neutrinos come primarily from $K$'s, this effect is
small for high energy neutrinos.
A bigger effect at low energy is the solar cycle 
variation,
which occurs because the magnetic field of the solar wind prevents
many low energy cosmic rays from reaching the earth's location in
the solar system.
\par If all neutrinos came from the same height in the atmosphere,
there would be a strict relationship between the local neutrino zenith
angle in an underground detector and the neutrino pathlength $L$ which
is relevant for neutrino oscillation analysis.  
Those production height 
distributions vary with both neutrino angle and energy.  A useful
parameterization has been performed by Ruddick\cite{ruddick}.
A comparison of his parameterization with 
a full Monte Carlo at two choices of
energy and angle are shown in Figure 1.

\begin{figure}[htbp]
\vspace{9pt}
\includegraphics*[width=3in]{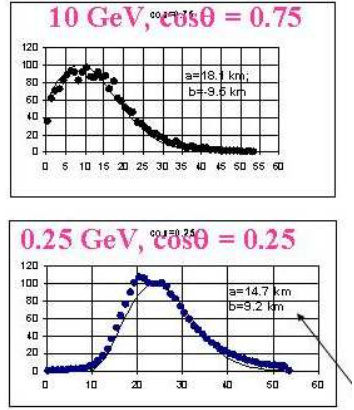}
\caption{Production height distributions for two neutrino energies
and angles.}
\label{fig:ruddick}
\vspace*{-12pt}
\end{figure}

\section{Detection of Atmospheric $\nu$}
\par Two classes of atmospheric neutrino events are considered in
this paper.  A neutrino may interact in an underground detector, or
it may interact outside the detector and make a muon which is detected.
Events of the first category include 
the contained events and partially contained events of Soudan~2, IMB
and Kamiokande and the semi-contained events of MACRO.  
Events of the second
category include the upward throughgoing $\mu$'s and stopping $\mu$'s of
MACRO and Kamiokande and the horizontal $\mu$'s of KGF, Soudan~2 and
Frejus.  An attempt to catalog the world total of atmospheric $\nu$'s
is given in Table 1.  Only neutrino candidates which are background free
or background subtracted have been included.  It is clear that
Super-Kamiokande dominates the total.
\begin{table}[htbp]
\begin{tabular}{|l|r|r|}  \hline
Experiment & Contained & $\nu$ induced $\mu$  \\ \hline
CWI/SAND\cite{sand} & 0      & 121  \\
KGF\cite{kgf}      & 100    & 229  \\
NUSEX\cite{nusex}    & 40     &   0 \\
Soudan 1\cite{soudan1} &  1     &   0 \\
Frejus\cite{frejus}   &  271   &  44 \\
IMB\cite{imb}      & 935    & 624  \\
Kamiokande\cite{kamiokande}  & 557 &    372  \\
Soudan~2\cite{soudan2} & 561 & 73  \\
LVD* & 0 & ?  \\ 
Baksan*\cite{baksan} & 0 & 801  \\
MACRO\cite{macro} & 285 & 940  \\
AMANDA*\cite{amanda} & 0 & 204+  \\
BAIKAL*\cite{baikal} & 0 & 44+  \\
\hline\hline
Subtotal & 2750 & 3452 \\
\hline\hline
Super-Kamiokande\cite{sk} & 12785 &   $\sim$ 1850 \\
\hline
\end{tabular}
\caption{Atmospheric neutrino event totals.  The *'s represent
experiments that are still taking data.  The total, not
including Super-Kamiokande is 6202 $\nu$ events}
\end{table}
\section{History}
\par
\begin{figure}[htbp]
\vspace{9pt}
\includegraphics*[width=3in]{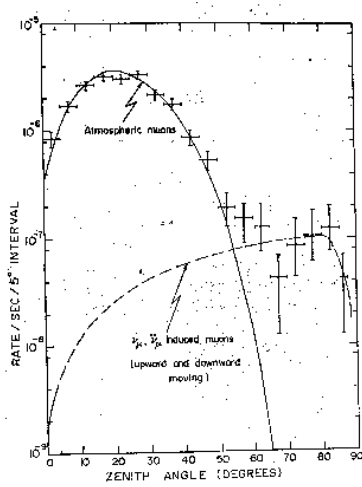}
\caption{Zenith angle distribution of muons in KGF.}
\label{fig:zenkgf}
\vspace*{-12pt}
\end{figure}

 The first reported atmospheric $\nu$ was measured in the KGF
experiment\cite{kgf} using a set of telescope counters consisting of
 iron, flash tubes and scintillator.  That telescope was 
operated on several levels of the KGF mine, but started at
7600 MWE level where the nucleon decay and monopole detectors 
were later built.  The measured angular distribution in the latter
detectors shows a clear separation between atmospheric $\mu$'s
and $\nu$ induced $\mu$'s.  This is shown in Figure \ref{fig:zenkgf}.
Eventually, 100 contained events and 229
$\nu$ induced horizontal $\mu$'s were measured at KGF.  
\par The first recorded atmospheric $\nu$ was measured on 23 
February, 1965 in the South African Neutrino Detector (CWI/SAND)
built by the Case-Western-Irvine Group.\cite{sand}
Again, the signature for an atmospheric $\nu$ was the projected
zenith angle.  Due to the extreme depth of that detector (8890 MWE), the
zenith angle where an atmospheric muon and neutrino induced muon can
be clearly separated was about 50$^o$.
\begin{figure}[htb]
\includegraphics*[width=2.5in]{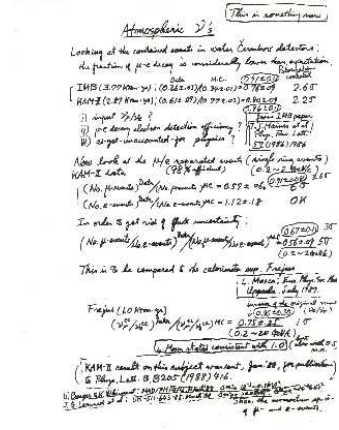}
\caption{Koshiba's slide from 1988 Fermilab Workshop.}
\label{fig:kosh}
\vspace*{-12pt}
\end{figure}

\par More atmospheric neutrino data became available in the
1980's from large detectors which were built to search for
nucleon decay.  The first hint of the atmospheric neutrino problem,
which was also known as the ``too few nu mu" problem came in the
IMB1 data.  They measured the number of delayed coincidences due 
to muon decays.\cite{haines}  This distinguished 
$\nu_\mu n \rightarrow \mu^- p$ 
and 
$\bar{\nu}_\mu p \rightarrow \mu^+ n$ 
followed by
$\mu \rightarrow e \nu \bar{\nu}$ from 
$\nu_e n \rightarrow e^- p$ 
and 
$\bar{\nu}_e p \rightarrow e^+ n$ 
events.  They 
reported, ``The simulation predicts 34\% $\pm$ 1\% of the events
should have an identified muon decay while our data has 26\% $\pm$ 
3\%.  This discrepancy could be a statistical fluctuation or a
systematic error due to (i) an incorrect assumption as to the ratio
of muon $\nu$'s to electron $\nu$'s in the atmospheric fluxes, (ii)
an incorrect estimate of the efficiency for our observing a muon
decay, or (iii) some other as-yet-unaccounted-for physics."  It
has turned out to be the latter.  
\par To learn from an anomaly, there needs to be
not only results inconsistent with expectation, but also someone to 
take it seriously and to carefully study alternative explanations.  The 
first detailed evaluation of the atmospheric neutrino problem that I
personally saw was  a discussion by Koshiba at the 1988 New Directions
in Neutrino Physics Fermilab Workshop.  There he showed the
transparency in Figure \ref{fig:kosh}.  He 
compared the ``ratio-of-ratios" defined as
\begin{equation}
R = \frac{ (\nu_\mu/\nu_e)^{data} }   
         { (\nu_\mu/\nu_e)^{MC}   }
\end{equation}
 for several experiments and several different methods.
The Water Cerenkov Detectors could measure the ratio two different
ways, with muon decays and with ring fits.  The initial measurements
of the iron calorimeters, Frejus and NUSEX got higher values consistent
with unity, but NUSEX had quite low statistics, and even Frejus had
a noticeable $\nu_\mu$ deficit if one looked only at the fully
contained events.  The period 1988-1995 was the period of the
``ratio-of-ratios" or the ``atmospheric neutrino anomaly".  Several
attempts to understand the anomaly either as a systematic effect or
as an error in the atmospheric neutrino Monte Carlos did not succeed.
\par In 1994, the Kamiokande experiment\cite{kamiokande} showed 
another data set, their
Multi-GeV data.  In the Multi-GeV data, not only was there a deficit
of $\nu_\mu$, and a value of $R$ below unity, but there was also a 
zenith angle distribution consistent with neutrino
oscillations.  By contrast,
their zenith angular distribution for the Sub-GeV contained events was
flat.   This suggested a higher value of
$\dmsq$ than currently seen.
It has been widely noticed that the parameter space plots ( in
$\mxang$ and $\dmsq$) for Kamiokande and Super-Kamiokande did not overlap.
It is interesting to compare the zenith 
angle plots of Kamiokande data and
Super-Kamiokande data for Sub-GeV and Multi-GeV, e and $\mu$.  In
such a comparison, no data points disagree by much more than one sigma.
The disagreement in parameter space comes about in the fits.  It is
worthwhile to point out that fits for $\Delta m^2$ are not gaussian,
and that it is possible to have multiple solutions
similar to the degeneracies now
facing experiments planning to measure $\theta_{13}$ and $\delta_{CP}$.

\section{Gedanken History}
\par Let's imagine the status of atmospheric neutrinos for
two scenarios in which 
history had been different.  

\par The largest 
underground experiments
for nucleon decay and magnetic monopole detection were
motivated by Grand Unified Theories.  In the absence of such motivation,
the only experiments which measured atmospheric $\nu$'s would have
been CWI/SAND, Baksan and LVD.  
Solar neutrino experiments would have proceeded
on their same time scale (or even faster).  There would likely have
been a greater interest in the reported neutrino oscillation
signature from the LSND experiment\cite{LSND}, as well as in proposed
short and intermediate baseline experiments such as BNL to Long Island,
CERN to JURA and the never-realized Fermilab COSMOS experiment.  The
solar neutrino results in the 80's and 90's 
might have motivated some forward thinking
individuals to propose a large underground experiment to measure
atmospheric neutrinos.  There would probably have not been great
enthusiasm for such an experiment, most sensitive to large mixing angles,
until after the SNO results in 2002.  Even then, I think it would have
been a hard project to realize.
\par History would also have been quite different if the Super-Kamiokande
accident had happened during its first fill in 1995.   Confidence
in rebuilding the detector might not have been possible in the absence
of the 5 successful years of running the detector.  The K2K run would
not have taken place.  MINOS, which had already been approved, would
have continued to plan with its high energy beam, most sensitive to
larger values of $\Delta m^2$.  New data from MACRO and Soudan~2 would
be tending to support lower values of $\Delta m^2$, but given the
non-gaussian (and hence unintuitive) nature of $\Delta m^2$ fits, the
situation would be fairly confused, which would have lead skeptics 
to doubt conclusions about atmospheric
neutrino oscillations.  The latest Soudan~2 and MACRO data analyses would 
be quite relevant in trying to sort out the situation.

\section{Recent Results from Baksan}
\par The Baksan Underground Scintillator Telescope has been taking data since
the 1980's, with its four layers of scintillator detectors under 
a mountain at a minimum 850MWE.  At Neutrino 2000, they reported
on data from December 1978 to January 2000, corresponding to 15.7
years of livetime.  They measured 801 upward $\mu$'s with an
expected rate (in the absence of oscillations) of 941.6.  The zenith
angle distribution was not in close agreement with oscillation fits, 
though
the rates agree with other experiments.  They now have
about another 10\% increase in statistics, but will not present a
new analysis until next year or later, when the increase is 25\% or more.

\begin{figure}[htbp]
\vspace{9pt}
\includegraphics*[width=3in]{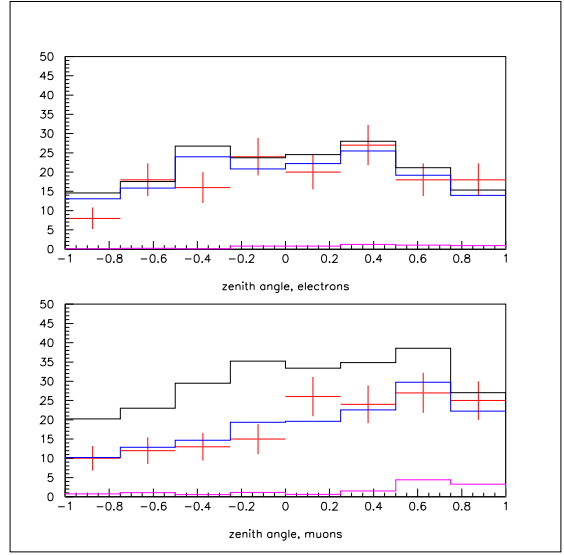}
\caption{Zenith angle distributions in Soudan~2 for e (top) and
$\mu$ (bottom).  Shown are data points, and histograms with and
without oscillation fits.  A small background is also shown.}
\label{fig:zensoud}
\vspace*{-12pt}
\end{figure}

\section{Recent Results from Soudan~2}
\par
\begin{figure}[tbh]
\vspace{9pt}
\includegraphics*[width=3in]{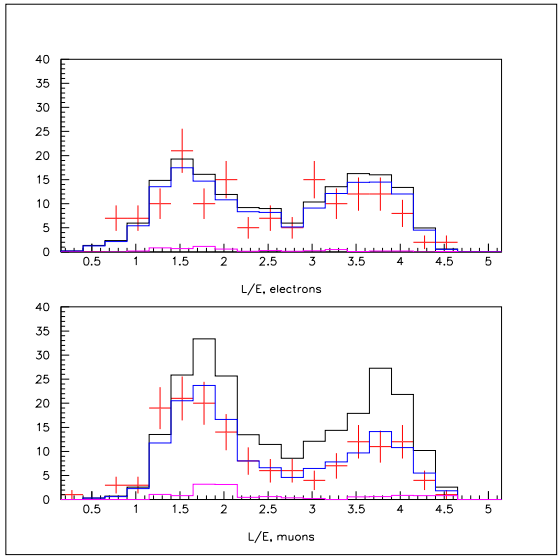}
\caption{$\log_{10}$ $L/E$ distribution in Soudan~2 for e(top) and
$\mu$(bottom).  The histograms have the same meaning as in Figure 4.}
\label{fig:loesoud}
\vspace*{-12pt}
\end{figure}

 Soudan~2 is a very fine-grained drifting calorimeter 
with drift cells located in a honeycomb pattern
of iron plates.
Since Neutrino 2000, the Soudan~2 experiment has finished taking
data.  It finished in July 2001 with a total 5.91 fiducial kt-year for
contained event data.  Details of the current status of Soudan~2
analyses for atmospheric neutrinos can be found in Reference
\cite{sanchez}.   A recent feature of the Soudan~2 data analysis
is that a 15\% problem with the electron energy scale has been resolved.
Data analysis now includes the partially contained events, and parameter
space fitting is done with a Feldman-Cousins\cite{feldman} type
of analysis.  A feature of Soudan~2 which has been exploited for
some time is the ability to measure recoil protons, which appear
in this very fine-grained detector as short, straight, heavily
ionizing tracks from the main vertex.  The events which have an
identified recoil proton allow the neutrino zenith angle, and hence
the $L$ relevant for $L/E$ analyses, to be reconstructed with much
greater accuracy.   Soudan~2 defines its high-resolution sample to
be the high energy quasi-elastics, the low energy quasielastics
with a recoil proton and the high energy multiprongs.  
The zenith angle distributions of
the electrons and muons are shown in Figure \ref{fig:zensoud}
and the $L/E$ distribution in Figure \ref{fig:loesoud}.  

\par
\begin{figure}[bht]
\vspace{9pt}
\includegraphics*[width=3in]{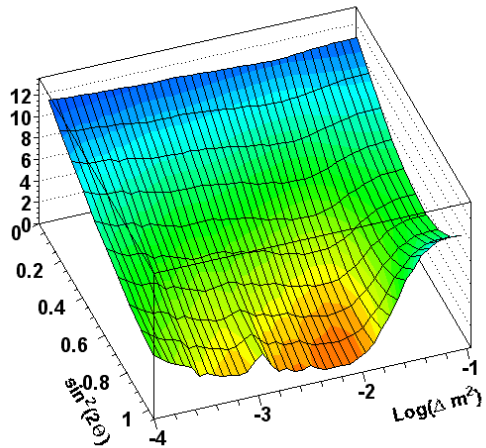}
\caption{Likelihood surfaces for Soudan 2 data.}
\label{fig:parsoud}
\vspace*{-12pt}
\end{figure}

 The Feldman-Cousins analysis involves calculating the likelihood
of the data as function of $\mxang$ and $\dmsq$ parameters
for the data and for Monte Carlo data.
A confidence level diagram is calculated from
the difference in log-likelihood.  This is shown in Figure 
\ref{fig:parsoud}.  The best fit occurs at a value of 
$\dmsq$ = 0.010 $eV^2$
and $\mxang$ = 0.97.  It should be noted that the valley near
the best fit is quite flat, particularly towards lower values of
$\dmsq$.  

\section{Recent Results from MACRO}
\par The MACRO experiment is a large area detector in the Gran Sasso
laboratory, consisting of 
3 towers of scintillation counter and 14 horizontal planes of 
streamer tubes.  MACRO completed its data taking in the last two
years, finishing with 5.52 live years in December 2000.  With
excellent timing, MACRO was able to separate upward and downward
going events, but the trigger gave poor acceptance for horizontal
muons.  The categories of events in MACRO were 809 + 54 [background 
subtracted + background] (1122 [Monte Carlo]) upthroughgoing muons,
154 +7 (285) internal upgoing (IU) events, and 262+10 (375) upgoing
stopping muons (UGS) plus internal downgoing muons (ID).
\par The shapes of the zenith angle distributions 
of the MACRO event samples are sensitive to neutrino oscillations.
For the throughgoing muons, the ratio of events with $-1 < \cos \theta
< -0.7$ to events with  $-0.4 < \cos \theta < 0 $ can be used to
distinguish $\nmt$ from $\nms$ oscillations because of a matter 
effect\cite{montaruli}.  The $\nu_\tau$ has a neutral current 
interaction in matter while the $\nu_s$ does not, and this
difference in interaction leads to a different expected
oscillation probability.  Based on the ratio test, $\nms$
oscillation has only a 0.033\% probability of fitting the data,
and is disfavored by $>99\%$CL compared to the best fit $\nmt$
oscillation.
\begin{figure}[thb]
\vspace{9pt}
\includegraphics*[width=2.8in]{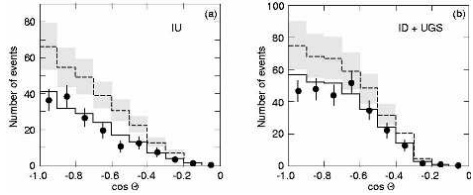}
\caption{Angular Distributions of MACRO low energy neutrino data,
along with expected shapes with and without oscillations.  The
shaded area shows the uncertainty of the flux estimate.}
\label{fig:zenmacro}
\vspace*{-12pt}
\end{figure}

\par The angular distribution of the MACRO low energy data is given
in Figure \ref{fig:zenmacro}.  For both the IU and UGS+ID data sets,
the data falls below the expectation (without oscillations),
 even taking into account the
normalization uncertainty of the expectation.  And together with
the throughgoing data, they form an acceptable fit to the same
neutrino oscillation parameters.  
\par
\begin{figure}[tbh]
\vspace{9pt}
\includegraphics*[width=3in]{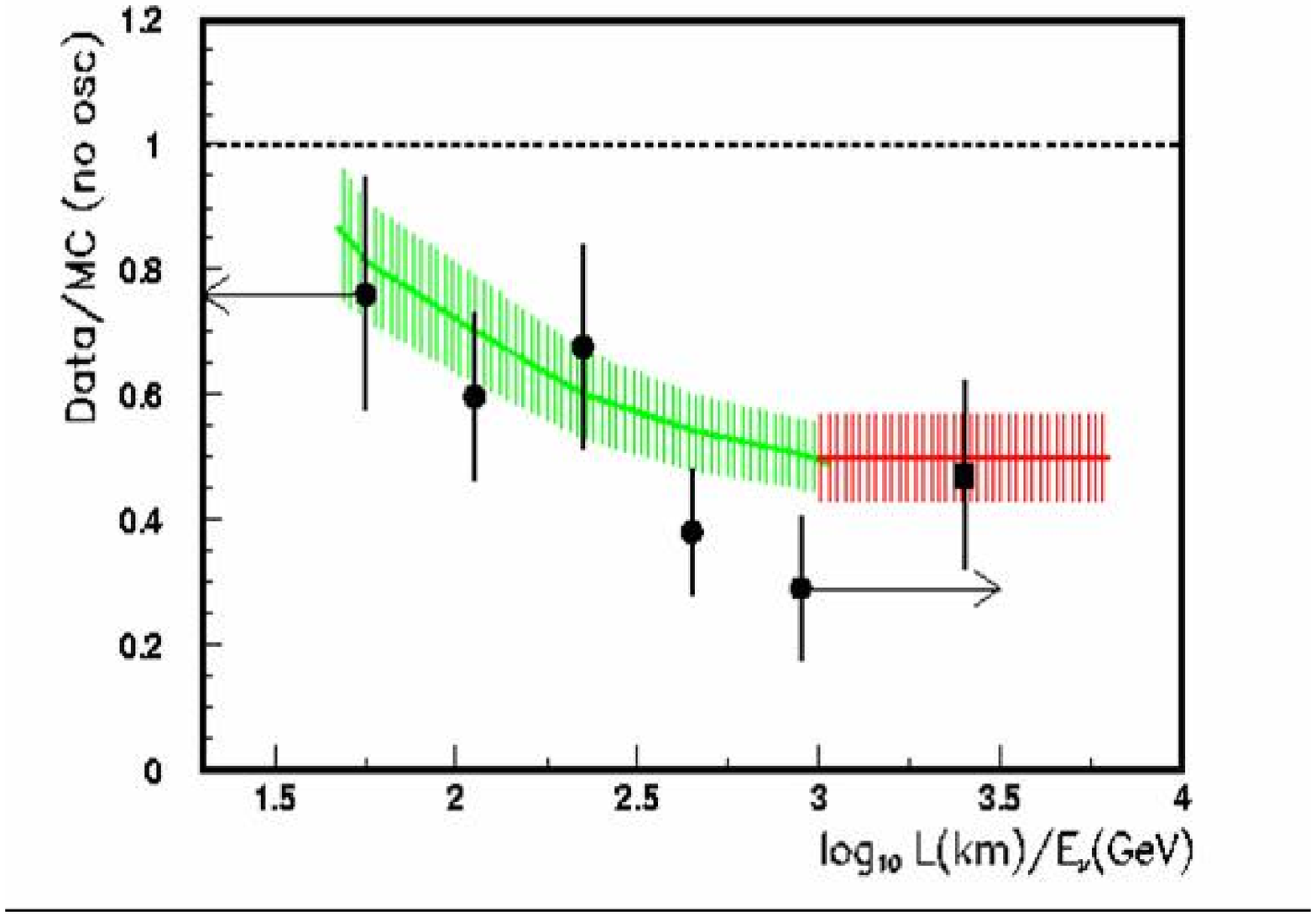}
\caption{$L/E$ distribution of MACRO data based on multiple
scattering estimation.}
\label{fig:loemacro}
\vspace*{-12pt}
\end{figure}

 MACRO has also recently shown new results which use muon energy
in the analysis.\cite{am2}  The muon energy is not measured calorimetrically or
magnetically.  However, the multiple scattering in the detector
is dependent on the energy on average in a known way.  The projected
displacement from a straight line fit for a relativistic muon
is inversely proportional
to the muon momentum.  They have
obtained an improvement of the space resolution using the limited
streamer tubes in drift mode, and through an analysis of the multiple
scattering, obtained an energy estimate on a muon by muon basis.
With that energy, and the zenith angle which gives an estimate of the
neutrino flight path, they have made an $L/E_\nu$ distribution.  This
is shown in Figure \ref{fig:loemacro} along with the expected distribution
for the best fit oscillation parameters.
\par
\begin{figure}[htbp]
\vspace{9pt}
\includegraphics*[width=2.5in]{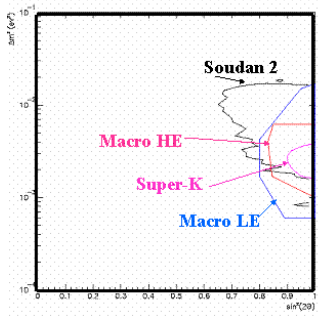}
\caption{Comparison of parameter space estimates for MACRO and Soudan~2}
\label{fig:param}
\vspace*{-12pt}
\end{figure}

 The allowed parameters ($\dmsq$, $\mxang$) according to the
low energy and high energy MACRO analyses are shown in Figure 
\ref{fig:param} along with the analysis of Soudan~2.  The results of
both experiments agree with each other and with 
Super-Kamiokande.  Each experiment also rejects the null hypothesis of
no oscillations with high probability.


\begin{figure}[tbh]
\vspace{9pt}
\includegraphics*[width=3.5in]{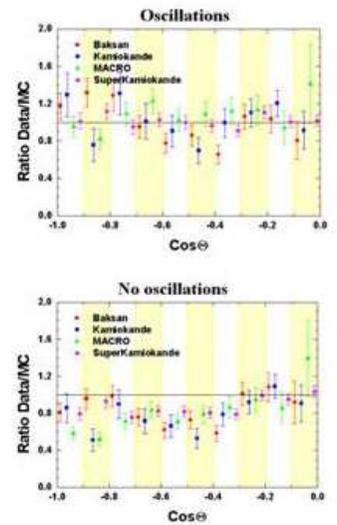}
\vspace*{-32pt}
\caption{Comparison of zenith angle distributions of throughgoing muons}
\label{fig:msum}
\vspace*{-22pt}
\end{figure}

\section{Summary}
\par There are several experiments besides Super-Kamiokande which 
collectively have observed atmospheric neutrinos.  Analyses of these
data agree with the essential Super-Kamiokande result:
neutrino oscillations are required to account for the data.  Taken as a
whole, contained event data from IMB, Kamiokande, Soudan~2, and Frejus
all agree with a 30\%-40\% deficit of $\nu_\mu$ induced events, and also
agree with the higher statistics 
Super-Kamiokande
results.   New features from Soudan~2 data strengthens that conclusion.
Soudan~2 now has an up/down difference in its contained events which
is statistically significant, and supports the region of parameter space
found by Super-Kamiokande.  With its recoil proton identification and good 
angular resolution, Soudan~2 has the resolution, but not the statistics,
to see the ``reappearance" in the $L/E$ distribution.
\par Detailed analyses of $\nu$ induced muons support a similar conclusion.
All experiments see a deficit of $\nu_\mu$ consistent with the 
Super-Kamiokande observations, and an angular distribution which is 
more consistent with the neutrino oscillation hypothesis than with
the null hypothesis, as shown in Figure \ref{fig:msum}, a comparison 
of data from Baksan, Kamiokande, MACRO and Super-Kamiokande.\cite{mikhaev}
It is interesting to point out that the probability of the best fit
for all four experiments is less than 30\% each, so it is worthwhile to
continue to investigate whether some aspect of the physics is not being modeled
correctly.
\par It is left to another speaker at Neutrino 2002 to consider 
the proposals for future atmospheric neutrino oscillation 
projects.\cite{fatis}  However, I want to make a general observation,
that much of the progress in atmospheric neutrinos has come
from experiments that were designed to search for GUT predicted
nucleon decay.  The motivation for much larger experiments to
continue that search is quite strong, and we as a field have the
technical means to accomplish such a search in our lifetime.  I
hope that search takes place.  If it does, a much firmer understanding
of atmospheric neutrinos will be an
inevitable outgrowth of such an effort.

\end{document}